\documentclass[twocolumn]{article}
\usepackage[margin=1in]{geometry}

\usepackage{amsmath,amsfonts}
\usepackage{algorithmic}
\usepackage{algorithm}
\usepackage{array}
\usepackage[caption=false,font=normalsize,labelfont=sf,textfont=sf]{subfig}
\usepackage{textcomp}
\usepackage{stfloats}
\usepackage{url}
\usepackage{verbatim}
\usepackage{graphicx}
\usepackage{cite}
\usepackage[export]{adjustbox}

\usepackage{atbegshi}

\AtBeginShipout{\AtBeginShipoutUpperLeft{%
  \put(\dimexpr1in+\oddsidemargin\relax, -\dimexpr1in+\topmargin+\headheight+\headsep+\textheight+\footskip-8
pt\relax){%
    \makebox[\textwidth][c]{\fontsize{7}{12}\selectfont Approved for Public Release; Distribution Unlimited: AFRL-2023-2308, 12 May 2023}
  }%
}}

\usepackage[T1]{fontenc}

\usepackage{times}
\usepackage{pifont}

\usepackage{ulem}
 
\usepackage{hyperref}
\usepackage{wrapfig}
\usepackage{eucal}

\usepackage{colortbl} 
\usepackage{enumerate}

\newcommand*{\trim}[1]{%
  \trim@spaces@noexp{#1}%
}

\newcolumntype{P}[1]{>{\centering\arraybackslash}p{#1}}
\newcolumntype{R}[1]{>{\raggedleft\arraybackslash}p{#1}}                                                                                                                                               
\newcolumntype{L}[1]{>{\raggedright\arraybackslash}p{#1}}  
\usepackage{comment}

\usepackage{bm}

\begin{document}

\title{A Review of the Duality of Adversarial Learning in Network Intrusion: Attacks and Countermeasures}

% 
% authors 
  \author{
  Shalini Saini\\
  \texttt{Texas A\&M University - Central Texas}
  \and
  Anitha Chennamaneni \\
  \texttt{Texas A\&M University - Central Texas}
  \and
  Babatunde Sawyerr \\
  \texttt{University of Lagos}
}
\date{}
\maketitle

\begin{abstract}
Deep learning solutions are instrumental in cybersecurity, harnessing their ability to analyze vast datasets, identify complex patterns, and detect anomalies. However, malevolent actors can exploit these capabilities to orchestrate sophisticated attacks, posing significant challenges to defenders and traditional security measures. Adversarial attacks, particularly those targeting vulnerabilities in deep learning models, present a nuanced and substantial threat to cybersecurity. Our study delves into adversarial learning threats such as Data Poisoning, Test Time Evasion, and Reverse Engineering, specifically impacting Network Intrusion Detection Systems. Our research explores the intricacies and countermeasures of attacks to deepen understanding of network security challenges amidst adversarial threats. In our study, we present insights into the dynamic realm of adversarial learning and its implications for network intrusion. The intersection of adversarial attacks and defenses within network traffic data, coupled with advances in machine learning and deep learning techniques, represents a relatively underexplored domain. Our research lays the groundwork for strengthening defense mechanisms to address the potential breaches in network security and privacy posed by adversarial attacks. Through our in-depth analysis, we identify domain-specific research gaps, such as the scarcity of real-life attack data and the evaluation of AI-based solutions for network traffic. Our focus on these challenges aims to stimulate future research efforts toward the development of resilient network defense strategies.

\end{abstract}

\renewcommand{\thefootnote}{}  % Remove footnote numbering
\footnotetext{This research was supported in part by the Air Force Research Laboratory (AFRL) under award FA8750-23-C-0085.}
\renewcommand{\thefootnote}{\arabic{footnote}}  % Restore default footnote numbering

\label{sec:intro}
\section{Introduction}
Cybersecurity is critical in today's interconnected world, where cyber threats, such as malware, phishing, ransomware, and data breaches, continue to evolve and pose significant risks to individuals, organizations, and even national security \cite{cyber-Tulsa, sunburst-hack}. The integration of embedded systems has revolutionized critical sectors like Healthcare, Automated Vehicles, and National Defense, relying on advanced technology and data integrity for improved efficiency, safety, and security. Nonetheless, the integration has also exposed network components more susceptible to cyber attacks and restricting their practical and secure implementation \cite{saini2022predatory, young2020towards}. Current trends indicate that approximately 67\% of the world's population has access to the internet, with social media platforms playing a significant role in this accessibility \cite{statista2024}. However, this increased connectivity has also expanded the attack surface for adversaries, resulting in heightened cybersecurity threats. Notably, the economic impact of cyber attacks has grown substantially, with a global estimated loss of approximately \$6 trillion in 2021, doubling the costs recorded in 2015 \cite{cyber-Tulsa}. As a result of a 2022 security breach at CommonSpirit Health, the personal data of 623,774 patients was compromised. During the incident, one of the affected hospitals experienced a computer system outage, leading to a reported case of a 3-year-old patient receiving a significant overdose of pain medicine \cite{ransomware-2022}. Worldwide end-user spending on security and risk management is projected to total \$215 billion in 2024, an increase of 14.3\% from 2023 \cite{gartner2023}. 

With rapid computational advancements, the abundance of big data, and the application of neural networks, machine learning has emerged as a vital tool for shaping contemporary defense strategies. Despite this significance, ML/DL-based solutions face susceptibility to adversarial learning attacks, challenging the efficacy of current defense protocols. As a result, a continuous effort is needed to formulate defensive strategies that can effectively adapt to the dynamic landscape of such threats. The development of counter-defenses that can effectively adapt to the evolving nature of these attacks remains an ongoing challenge \cite{khan2022deep}. Through our evaluation, we identify the current research gaps in adversarial learning pertaining to Network Intrusion Detection Systems (NIDS). Our contribution lies in providing a baseline understanding of the existing research breadth and presenting the potential future directions in the field of ML/DL-based NIDS. The objective is to facilitate the development of robust ML/DL-based NIDS that effectively harness evolving technological capabilities, while also demonstrating resilience against both known and emerging security threats.
\vspace{2mm} 

\noindent\textbf{Research Objectives and Contributions}

Our work is motivated by the growing need to understand the impact of artificial intelligence methodologies and technological advancements on innovative research and their specific implications in real-world applications within the field of cybersecurity. This work is aligned to keep pace with the evolving landscape of cybersecurity and leverage the emerging technologies by addressing following research objectives. 
\begin{itemize}
    \item Conduct a comprehensive review of adversarial learning-based attacks and defenses within cybersecurity, with a specific focus on NIDS.
    \item Identify and analyze the state-of-the-art ML/DL techniques and methodologies employed in adversarial attacks against NIDS within varying attacker’s knowledge.
    \item Gain insights into the current state of research on data poisoning, test-time evasion, and reverse engineering attacks targeting NIDS.
    \item Examine the existing defense mechanisms and countermeasures against adversarial attacks in NIDS, highlighting their limitations and scope for improvement.
\end{itemize}

By addressing our research objectives related to adversarial attacks and defenses in NIDS, our work makes significant contributions to the field of cybersecurity. The key contributions of our research include:
\begin{itemize}
    \item A comprehensive analysis of the existing literature, our study provides valuable insights into adversarial learning-based attacks and defenses in NIDS.
    \item Our work identifies the critical research gaps of practical implementation of advanced ML/DL methodologies in NIDS addressing the real-world challenges.
    \item We present an in-depth investigation into data poisoning, test-time evasion, and reverse engineering attacks targeting NIDS, uncovering emerging trends, novel techniques, and potential future directions in the field.
    \item Our work summarizes the current state of research and provides a solid foundation for advancements through developing innovative defense mechanisms for implementing resilient defense strategies in real-world NIDS deployments.
\end{itemize}
%----Figure 1----------------------------------
\begin{figure}
\centering
\includegraphics[width=0.5\textwidth, right]{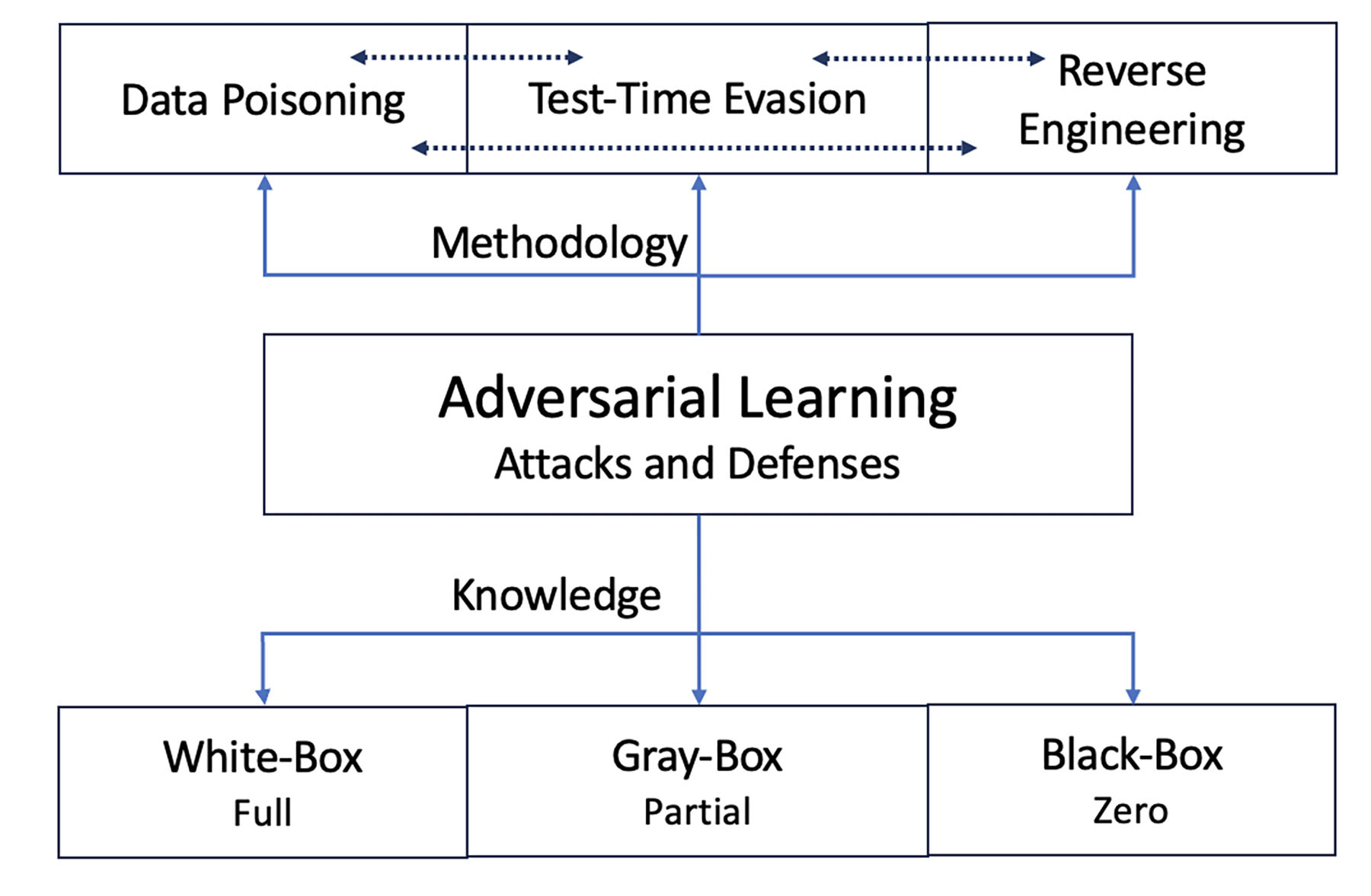}
\caption{Data Poisoning, Test-Time Evasion, and Reverse Engineering Adversarial Learning Attacks and Defenses for ML/DL-Based NIDS. }
\label{fig:focus}
\end{figure}
As presented in Figure \ref{fig:focus}, we explore ML/DL-based adversarial attacks and defenses within the context of Data Poisoning, Test Time Evasion, and Reverse Engineering. We examine the various threats, attacks, and possible defenses in which both attackers and defenders may possess different levels of knowledge, such as White-Box (Full), Black-Box (Zero), and Gray-Box (Partial) understanding of the model, algorithm, and training data. As Figure \ref{fig:focus} depicts, data poisoning can be instrumental in test time evasion attacks. Similarly, reverse engineering can provide significant details to sharpen the attacks. At the same time, by analyzing the model's vulnerabilities and threats to data integrity, reverse engineering can be pivotal to develop robust defense mechanisms to minimize the damage by potential attacks.

Rest of the paper is structured as follows: Section II presents the research methodology, while Section III provides a comprehensive background on network intrusion, including a taxonomy of network attack types and an overview of ML/DL methods applicable in network intrusion domain. In Section IV, we outline existing work in the field and highlight the unique aspects of our study. Section V focuses on adversarial learning attacks and defenses with benchmark network datasets commonly employed in research studies. Moreover, within Section V, we specifically delve into the areas of data poisoning, test-time evasion, and reverse engineering attacks and defenses. Section VI discusses the limitations and challenges within the scope of our work. In Section VII, we present the discussion and future directions of our research, followed by the conclusion in Section VIII.

\label{sec:Methodology}
\section{Research Methodology}
In order to conduct a comprehensive search for research publications across multiple research repositories, we employ a strategic approach that involves the utilization of multiple keywords and phrases. By carefully selecting and combining relevant terms, we aim to cast a wide net and ensure the inclusion of diverse and pertinent literature. Our search strategy involves identifying key concepts and themes related to our research topic and constructing queries that encompass these facets. To ensure a thorough analysis, we consulted renowned research databases, including \textit{IEEE Xplore}, \textit{ACM Digital Library}, and \textit{SpringerLink}.  We also utilized \textit{Google Scholar} for the existing work in the field, additionally querying \textit{Base} and \textit{arXiv}. In addition to these databases, we leveraged \textit{Dimensions.ai} for its comprehensive research insights, allowing us to gather a broader perspective on the relevant literature \cite{dimensions.ai}. 
%----------------------------------------------
\begin{table*}
\centering
\caption{Key Search Queries in Defined Research Scope}
\label{tab:search}
\vspace{3mm}
\begin{tabular}{|L{5cm}|L{9.5cm}|}
\hline
\textbf{Search Scope}       & \textbf{Key Search Queries}                               \\ \hline
Adversarial Learning          & ML/DL Adversarial Learning                               \\ \hline
                              & Adversarial Learning in NIDS                             \\ \hline
Adversarial Learning Attacks  & Adversarial Learning Attacks in NIDS                     \\ \hline
                              & Data Poisoning Attacks                                   \\ \hline
                              & Backdoor Data Poisoning Attacks                          \\ \hline
                              & Test-Time Evasion Attacks                                \\ \hline
                              & Reverse Engineering Attacks                              \\ \hline
Adversarial Learning Defenses & Defense against adversarial attacks                      \\ \hline
                              & Defense against adversarial attacks in network intrusion \\ \hline
                              & Defenses for Data Poisoning adversarial attacks in NIDS  \\ \hline
                              & Defenses for Test-Time Evasion Attacks in NIDS           \\ \hline
                              & Defenses for Reverse Engineering Attacks in NIDS         \\ \hline
Attacker's Capabilities       & Black Box adversarial attacks and defenses in NIDS       \\ \hline
                              & White Box adversarial attacks and defenses in NIDS       \\ \hline
                              & Gray Box adversarial attacks and defenses in NIDS        \\ \hline
Attack Datasets               & NIDS Datasets                                            \\ \hline
                              & Network traffic attack datasets                          \\ \hline
                              & Benchmark network attack datasets                        \\ \hline
\end{tabular}
\end{table*}

% inclusion/exclusion criteria-----

\begin{table*}[!htb]
\centering
\caption{Research Literature- Inclusion and Exclusion Criteria}
\label{tab:inclusion-exclusion}
\vspace{3mm}
\begin{tabular}{|L{5cm}|L{10cm}|}
\hline
\textbf{Inclusion Criteria:}        & \textbf{Key Considerations}                                                                                         \\ \hline
Relevance to Adversarial Learning in NIDS &
  Papers directly address adversarial learning techniques in the   context of Network Intrusion Detection Systems \\ \hline
Focus on Adversarial Attacks and Defenses &
  Papers discuss adversarial attacks (e.g., data poisoning,   test-time evasion, reverse engineering) and defenses specific to NIDS \\ \hline
Methodological Alignment              & Papers focusing on machine learning, deep learning, and AI   methodologies in the context of NIDS                   \\ \hline
Specific Attack Types                 & Papers investigating data poisoning, backdoor attacks,   test-time evasion, and reverse engineering attacks in NIDS \\ \hline
Diversity of Defense Strategies       & Papers exploring various defense mechanisms against   adversarial attacks in NIDS                                   \\ \hline
Attacker's knowledge and Their Impact & Papers discussing black box, white box, and gray box   adversarial attacks and defenses in NIDS                     \\ \hline
Dataset Utilization                   & Papers utilizing NIDS datasets or proposing new datasets for   research purposes                                    \\ \hline
\textbf{Exclusion Criteria:}          &                                                                                                                     \\ \hline
Irrelevant Topics                     & Papers not directly related to adversarial learning in   NIDS                                                       \\ \hline
General Machine Learning or  Deep Learning Studies &
  Papers that do not specifically focus on NIDS or adversarial   learning in the context of network security \\ \hline
Non-Technical Papers                  & Non-technical papers, such as opinion pieces, editorials, or   reviews without novel research contributions         \\ \hline
Outdated Material                     & Papers published before a certain date (e.g., more than 5   years old) to focus on the most recent research         \\ \hline
Language and Publication Quality &
  Poorly written or low-quality papers, as well as papers not   published in reputable journals or conferences \\ \hline
\end{tabular}
\end{table*}
As per Table \ref{tab:search}, our base search was built around \textit{Adversarial Learning Attacks and Defenses in Network Intrusion Detection Systems}, and then we narrowed it down towards \textit{Data Poisoning}, \textit{Test-Time Evasion}, and \textit{Reverse Engineering}. We further categorized these topics into \textit{Black-Box}, \textit{White-Box}, and \textit{Gray-Box} adversarial learning attacks and defenses in NIDS. To achieve an extensive coverage of relevant publications in adversarial machine learning, a combination of the adversarial learning keyword with other appropriate terms such as perturbation, evasion, inference, model inversion, model stealing, and model poisoning  was employed. Additionally, concepts closely tied to network intrusion detection, including anomaly detection, cyberattack, and wireless and IoT networks, were taken into account to ensure a comprehensive scope. We executed around hundred initial AL-ML-DL-NIDS queries, and then we iterated and refined our search queries, incorporating synonyms, related terms, and variations to capture a broader range of relevant publications. We leveraged advanced search operators and filters provided by the research repositories to narrow down the results based on publication dates, authors, and other criteria. Through this meticulous process of applying inclusion and exclusion criteria as per Table \ref{tab:inclusion-exclusion}, we optimized the retrieval and compilation of research publications for this work, enabling us to gather a comprehensive and diverse collection of relevant studies from various sources.

\label{sec:background}
\section{Network Intrusion: Background and Taxonomy}
Machine learning has been applied in cybersecurity since the late 1990s and early 2000s. Early applications of machine learning focused on network intrusion detection and malware detection through file analysis and behavioral patterns. Over time, advanced machine learning techniques have found extensive applications in anomaly detection, threat intelligence, fraud detection, and spam filtering. Deep learning, a subset of machine learning, leverages neural networks to autonomously extract features and patterns from raw data, enhancing cybersecurity capabilities \cite{shone2018deep, ban2022malicious}. Deep learning is preferred over traditional machine learning due to its automatic feature extraction, hierarchical representation learning, scalability with large datasets, end-to-end learning capability, and flexibility across various data types \cite{shone2018deep, amutha2022secure}. However, these complex deep learning models are susceptible to adversarial attacks that can manipulate the data, model or both resulting in unintended output \cite{alatwi2021adversarial, he2023adversarial, xu2020adversarial, miller2020adversarial,corona2013adversarial}.

Signature-based detection and anomaly-based detection are two broad categories of network intrusion detection methods \cite{alatwi2021adversarial, javaid2016deep}. Signature-based Network Intrusion Detection Systems (SNIDS) detect intrusions by matching network traffic against pre-installed attack signatures. They are effective in detecting known attacks but struggles with unknown or new attacks \cite{kruegel2005reverse}. Anomaly-based Network Intrusion Detection Systems (ADNIDS) detect intrusions by identifying deviations from normal network patterns, without relying on pre-installed attack signatures. They are effective in detecting unknown or new attacks but may have higher false-positive rates \cite{javaid2016deep, nassif2021machine}. Elevated false-positive rates can result in alert fatigue, resource drain, operational disruption, loss of trust, compliance concerns, and increased costs, highlighting the importance of mitigating false positives to maintain system effectiveness, operational efficiency, and trustworthiness.

Deep learning approaches are used in NIDS to automatically learn relevant features, capture non-linear relationships, adapt to new threats, detect anomalies, and handle large-scale network traffic data for effective intrusion detection. Three major classes of deep learning architectures, namely the Generative (unsupervised), Discriminative (supervised) and Hybrid deep architecture provide a lot of flexibility to be useful and reliable in a wide range of problems \cite{lateef2019survey}.

\subsection{Network Attacks}
Network vulnerabilities refer to weaknesses or flaws in computer networks that can be exploited by malicious actors to gain unauthorized access, perform unauthorized actions, or disrupt network operations. While numerous defenses have been developed to identify attacks, the cycle of developing advanced attack techniques to surpass known defenses persists, creating an ongoing challenge to develop robust solutions. 

\noindent\textbf{Denial of Service (DoS) Attack} is a malicious attempt to disrupt a network, service, or website by overwhelming it with excessive traffic or exploiting system vulnerabilities to render the targeted resource unavailable. DoS attacks can be implemented as flooding attacks, protocol-based attacks, application layer attacks, and distributed DoS \cite{lai2023two, alrawashdeh2020defending}.

\noindent\textbf{Trojan} is malicious software or malware that disguises itself as a legitimate program or file to deceive users and gain unauthorized access. Trojans trick users into executing or installing them, often disguised as harmless or desirable, such as games, utility programs, or email attachments. Once a Trojan is executed, it can perform a variety of malicious activities without the user's knowledge or consent \cite{schwarzschild2021just, alrawashdeh2020defending, xiang2021reverse, xu2020defending, li2021deeppayload}.

\noindent\textbf{Ransomware} is malware that encrypts files or locks a computer, demanding a ransom for decryption or unlocking. It infects systems through email attachments, compromised websites, or software vulnerabilities. The victim's data becomes inaccessible until a ransom, often in cryptocurrency, is paid to the attackers. Encryption algorithms make it challenging to decrypt files without the decryption key \cite{ransomware-2022}.

\noindent\textbf{Botnet Attack} is a malicious operation where a network of compromised computers is utilized to carry out various harmful activities. The compromised computers (\textit{bots}) are infected with malware, enabling the attacker remotely. The botnet can launch DDoS attacks, spreading malware, exfiltrate sensitive data, or sending spam emails \cite{debicha2023tad, beigi2014towards}. 

\noindent\textbf{Brute Force Attack} is an unauthorized access method to systematically attempt all possible combinations of passwords or encryption keys until the correct one is discovered through trial and error to gain illicit access to systems or data \cite{tavallaee2009detailed}. 

\noindent\textbf{Heartbleed Attack} allows attackers to exploit a flaw in the OpenSSL implementation of the Transport Layer Security (TLS). By sending a maliciously crafted request to a vulnerable server, an attacker could trick the server into leaking sensitive information from its memory, including private keys, usernames, passwords, and other data \cite{CSE-CIC-IDS2018, sharafaldin2018toward}.

\noindent\textbf{Web Attack} is a malicious activity that targets vulnerabilities in web applications, servers, or infrastructure. It includes attacks like Cross-Site Scripting (XSS), SQL injection, Cross-Site Request Forgery (CSRF), Remote File Inclusion (RFI), DDoS, phishing, and brute force attacks \cite{kuppa2019black}.

\noindent\textbf{Infiltration Attack} is a network attack through exploiting vulnerabilities within the network or host machine. An attacker can gain access through exploiting vulnerabilities in software like Adobe Acrobat Reader to launch backdoors, malware, and other attacks \cite{gao2020backdoor, kuppa2019black}. 

\noindent\textbf{Zero-day exploits} are unknown threats that capitalize on undisclosed vulnerabilities, enabling attackers to employ novel attacking strategies before these vulnerabilities are identified or safeguarded against \cite{atwell2016reverse}.  These attacks can be leveraged swiftly by malicious actors to infiltrate systems, often evading conventional security measures and necessitating rapid response strategies.

\subsection{Deep Learning in Network Intrusion}
Deep learning techniques are commonly used in NIDS to analyze network traffic data and detect anomalies or intrusions. In NIDS, network traffic data is preprocessed and transformed into data representations suitable for deep learning models. This can involve extracting features from packet headers, flow-level information, or using raw payload data. 

Convolutional Neural Networks (CNNs) are effective as they can operate on network data represented as multidimensional arrays, such as time-series data or spectrograms. CNNs enable the learning of hierarchical representations of network traffic patterns, improving the ability to identify anomalies \cite{asnani2023reverse, liu2019mitigating}. Deep Neural Networks (DNNs) are more generalized and can process various types of data, including numerical, categorical, or textual data. They consist of multiple layers of interconnected neurons and are well-suited for tasks like regression, classification, and sequence-to-sequence mapping. DNNs are versatile in handling a wide range of machine learning problems, as they can learn complex representations and extract high-level features from the input data \cite{wang2019neural, wang2019not, papernot2017practical, liu2019mitigating, li2021deeppayload}. Recurrent Neural Networks (RNNs) are suitable for processing sequential data, making them applicable in NIDS for analyzing network traffic over time. RNNs can capture temporal dependencies and long-term patterns in network data, allowing them to detect anomalies based on deviations from learned temporal patterns \cite{amutha2022secure}.

Autoencoders and Variational Autoencoders (VAEs) are unsupervised deep learning models used in NIDS to learn efficient representations of normal network traffic. By training autoencoders on normal traffic, deviations or intrusions can be detected by measuring the reconstruction error \cite{mirsky2018kitsune, kuppa2019black}. Graph neural networks (GNNs) hold promise for network intrusion detection due to their ability to handle complex dependencies and relationships in network data \cite{wang2021intrusion}. Specialized deep learning architectures for NIDS that combine multiple layers to capture complex patterns and relationships in network traffic data, can enhance intrusions and anomaly detection capabilities \cite{zhang2020tiki}.

Transfer learning techniques have been applied in NIDS by utilizing pre-trained deep learning models. These models, originally trained on large-scale datasets like ImageNet, are fine-tuned on network traffic data to leverage the learned representations. This approach helps overcome the challenge of limited labeled data in NIDS and improves the overall detection performance \cite{chen2019stateful}. Federated Learning in NIDS revolutionizes network intrusion detection by enabling decentralized devices to collaboratively train models. This approach enhances detection accuracy and robustness without compromising data privacy or requiring data centralization. By leveraging the collective knowledge and utilizing insights from a multitude of devices, NIDS enhanced by federated learning effectively detects and combats network intrusions, emphasizing data security and privacy protection \cite{agrawal2022federated}.

\label{sec:related work}
\section{Related Work}
As technology evolves, cybersecurity threats in deep learning applications have advanced. Despite extensive research on NIDS attacks and defenses, a comprehensive exploration of Data Poisoning, Test-Time Evasion, and Reverse Engineering in this domain is lacking. Existing studies often address individual aspects, lacking a holistic overview. Initial adversarial learning vulnerabilities were examined within vision domain. However, it is critical to acknowledge domain-specific nuances when assessing both possible attacks and viable defenses. Rosenberg et al. highlighted the divergence between image-based adversarial learning attacks and defenses and those specific to cybersecurity, emphasizing the distinct nature of applied data features. Challenges unique to cybersecurity include seamlessly modifying network traffic without detection. While advocating for defense methods applicable to a wide array of attacks, the need for NIDS systems to swiftly and accurately detect specific threats in real-time complicates the implementation of robust defense mechanisms \cite{rosenberg2021adversarial}. 

Zhou et al. explored adversarial attacks and defenses in deep learning within the cybersecurity domain, focusing on Advanced Persistent Threats (APT). Their research centered on constructing a framework detailing a five-stage APT lifecycle for AL attacks and defenses. While they applied this framework to domains like image, video, audio, and text, the exclusion of network traffic data discussions limits its applicability to network intrusion \cite{zhou2022adversarial}. In another recent survey on NIDS adversarial attacks, He et al. emphasized critical issues arising from outdated benchmark datasets and the difficulty in generating authentic network features due to data instability. Moreover, the impracticality of modifying features for attacks lies in their inability to be transferable. They also discussed the scarcity of defense mechanisms against packet-level attacks \cite{he2023adversarial}. Jmila and Khedher studied shallow classifiers' efficacy in assessing robustness, enhancing them with Gaussian data augmentation. Their research suggested tailoring methods, parameters, and datasets specific to network intrusion detection scenarios, noting performance degradation with certain datasets and attack types \cite{jmila2022adversarial}. A survey by McCarthy et al. explored functionality-preserving adversarial attacks and defenses in the area of poisoning, evasion and transferability attacks in cybersecurity domain \cite{mccarthy2022functionality}. Another recent work by Vitorino et al. summarizes the state-of-the-art  approaches focusing on adversarial learning strategies to generating realistic adversarial examples to analyze and protect real-life scenarios. However, their work identified that more than 75\% of the reviewed research used known common methods and lacked novel strategies to generate adversarial samples \cite{vitorino2023sok}.

Ibitoye et al. categorized NIDS attacks into problem and feature space, with feature space representing data variables and problem space denoting input types like files or images. The study also mapped specific ML methods, such as supervised learning, to distinct ML task categories like classification \cite{ibitoye2019threat}. In a study, Apruzzese et al. revealed the limitations of utilizing deep learning methods for practical NIDS attacks. They discussed that feature-space attacks for NIDS are impractical in practice due to the difficulty of deriving adversarial traffic from features \cite{apruzzese2022modeling, han2021evaluating}. On applied datasets, approximately one third of the analyzed research work relied on the KDD99 dataset which has above 75\% duplicate records in test and train data \cite{sharafaldin2018toward, tavallaee2009detailed}. Other than the high reliance on outdated data sources, their analysis showed that only 10\% of the attacks studied are poisoning attacks, emphasizing the practical limitations of launching realistic NIDS attacks \cite{apruzzese2022modeling, sharafaldin2018toward}.

Our work differs from prior studies by providing a comprehensive literature review that systematically delves into the security and integrity aspects of adversarial learning attacks and defenses. Specifically, we focus on Data Poisoning, Test-Time Evasion, and Reverse Engineering within the context of ML/DL-based network intrusion solutions. Our work presents the most recent developments in the field of adversarial learning, specifically from 2018 to 2023-24. By primarily considering research conducted within this time-frame, we ensure that our work remains highly relevant and up-to-date. Moreover, we meticulously include both pioneering and milestone studies in the area of adversarial learning, allowing us to provide a comprehensive overview of the field's key advancements. Through our comprehensive analysis, our work fills a gap in the literature, providing valuable insights and paving the way for future advancements in this critical field of research.

\label{sec:Adversarial Learning Attacks}
\section{Adversarial Learning}
\begin{figure}[!htb]
\centering
\includegraphics[width=0.5\textwidth]{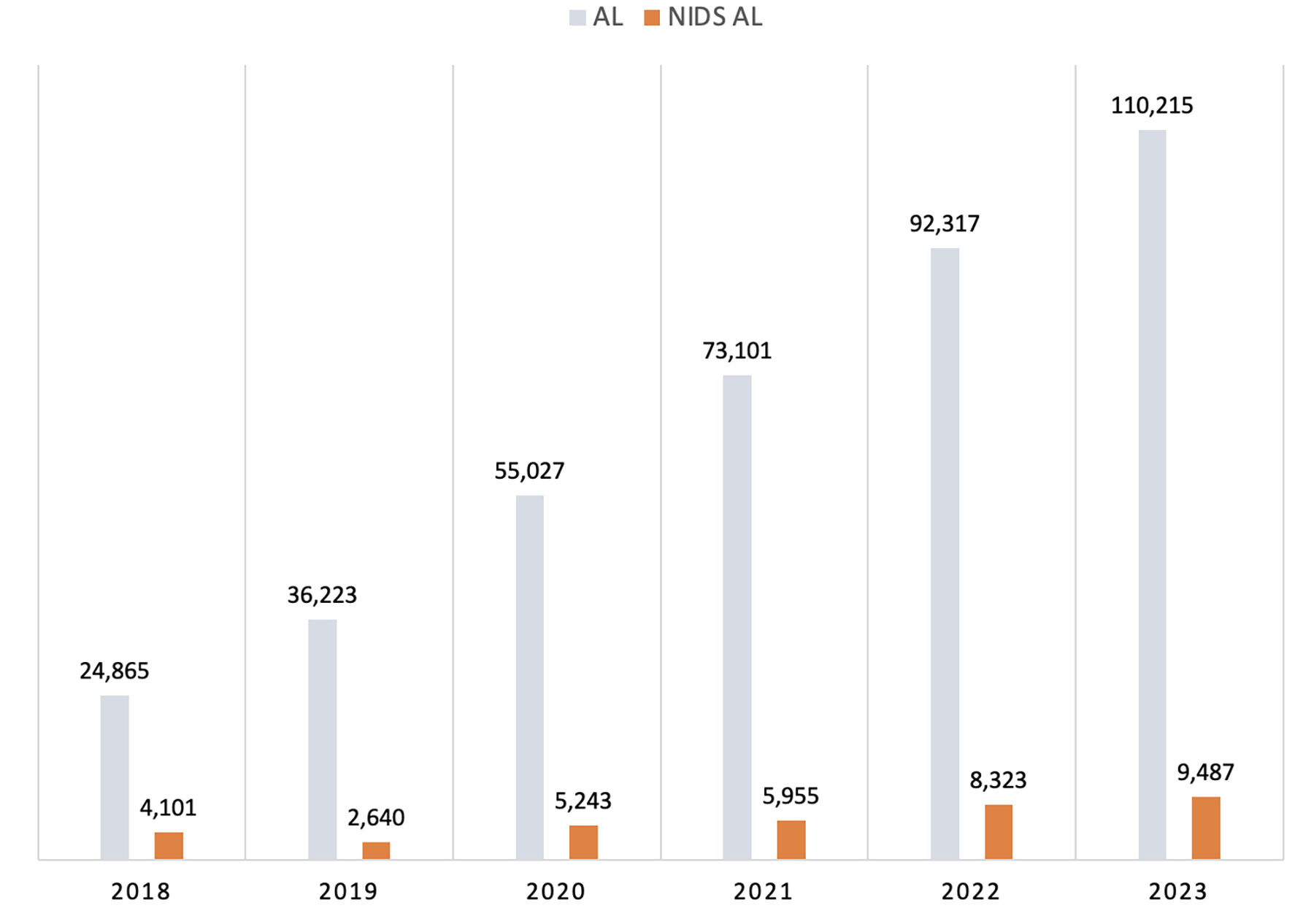}
\caption{2018-2023: Adversarial Learning Research Publications Trends}
\label{fig:al-trends}
\end{figure}

Approximately a decade ago, researchers identified adversarial learning as a significant vulnerability in neural networks. Adversarial learning pertains to their susceptibility to small, imperceptible changes in input data, as well as the manipulation of decision boundaries, which can lead to misclassifications \cite{szegedy2013intriguing, goodfellow2014explaining, athalye2018synthesizing}. In past few years, adversarial learning has grown as a field of great research interest because of its wide applicability and impact on artificial intelligence based solutions in different domains. As depicted in Figure \ref{fig:al-trends}, the field of adversarial learning has experienced a significant five-fold growth in the past five years. However, it is important to note that the existing literature predominantly concentrates on adversarial attacks and defenses in image, audio, and video domains. In contrast, research specifically dedicated to adversarial learning in the context of NIDS is limited, accounting for less than 10\% of all adversarial learning research\footnote{https://app.dimensions.ai/discover/publication}. While the existing literature may not be extensive, notable progress is evident in advancing knowledge within this domain. Concurrently, ongoing research in adversarial learning is driving a substantial transformation in the field of adversarial attacks and defenses for NIDS.  
\subsection{Adversarial Learning Attacks}
 
Adversarial learning attacks present a substantial challenge to the security and reliability of deep learning models. These attacks fall into different categories based on methodologies and approaches, which directly influence the strategies employed by attackers to create adversarial examples and exploit vulnerabilities \cite{zhang2022adversarial, zhou2022adversarial}. Targeted attacks aim to evade detection for specific malicious activities, while untargeted attacks disrupt the system by triggering false alarms or misclassifying normal traffic as malicious \cite{carlini2019evaluating, zhou2022adversarial}. Attacker's knowledge of model and the training data categorizes attacks as Black-Box (No knowledge), White-Box (Full knowledge), and Gray-Box (Partial knowledge) attacks. Other major classifications are based on network traffic manipulation strategies: feature-based, flow-based, and packet-based. 

Model manipulation attacks involve exploiting the transferability of adversarial attacks between models, corrupting the model, deceiving model through feedback mechanisms, and introducing malicious patterns during dynamic updates \cite{kim2021channel,heo2019fooling}. Availability of classified labeled network traffic data allows adversaries to evade accurate detection by manipulating traffic patterns, while physical layer attacks compromise infrastructure components \cite{sadeghzadeh2021adversarial}. 

Hybrid adversarial attacks on deep learning models involve combining multiple attack techniques to compromise model integrity and performance. These attacks can include a combination of evasion and poisoning techniques, as well as targeting strategies. By leveraging multiple strategies, adversaries create more sophisticated and challenging threats to deep learning models \cite{du2020hybrid}. Adaptive adversarial attacks continuously adapt to countermeasure updates and target NIDS feedback timing to compromise the system's learning process and effectiveness \cite{wang2019invisible}. Adversarial NIDS attacks include strategic manipulation of network traffic, subtle perturbations to evade detection algorithms, generative attacks that generate malicious traffic resembling benign patterns, and stealth attacks that hide malicious activities within legitimate traffic to avoid detection. \cite{shi2017evasion, warde201611, sun2021adversarial, tyukin2020adversarial, alatwi2021adversarial}.

\label{sec:Adversarial Learning Defenses}
\subsection{Adversarial Learning Defenses}
Adversarial learning defenses have advanced over the years,  encompassing techniques and strategies to fortify machine learning models against attacks. Approaches like adversarial training, detection/filtering, gradient masking, and randomized smoothing can provide effective defenses. Training-based defenses enhance adversarial robustness through adversarial training, robust feature extraction, model architecture modifications, compression and quantization, and diverse loss functions. These techniques enhance models against attacks by improving training data, resilient feature extraction, network architecture adjustments, model simplification, and optimization diversity \cite{bai2021recent, zhang2021robust}. Input transformation and reconstruction-based defenses enhance adversarial robustness by altering and reconstructing input data, guarding against perturbations and boosting model generalization. Ensemble-based defenses strengthen robustness by using multiple models, diversifying ensemble members through varied training to collectively improve defense efficacy against attacks \cite{tramer2017ensemble}.

Randomized Smoothing introduces random noise to input data during training and inference to enhance the model's decision boundaries and reduce vulnerability to adversarial perturbations \cite{cohen2019certified}. Carlini et al. proposed detailed evaluation methodologies with recommendations to address common flaws in defense evaluations \cite{carlini2019evaluating}. These include techniques such as gradient masking, which conceals the gradients utilized by attackers, robust optimization that enhances the classifier's capability to classify adversarial instances through parameter relearning, and adversarial example detection, a method that scrutinizes example distributions to differentiate and eliminate adversarial samples \cite{xu2020adversarial}.

\label{sec:NIDS Datasets}
\subsection{Network Intrusion Benchmark Datasets}

Accurate data interpretation is crucial in deep learning-based NIDS for representing normal and attack scenarios. While diverse attack datasets help identify and mitigate threats, adversaries exploit this knowledge to craft deceptive attacks. Challenges like limited availability, imbalanced class distribution, evolving techniques, and privacy concerns hinder the effectiveness of network attack datasets. Keeping datasets up-to-date becomes a challenge due to evolving techniques and the need for real attack instances \cite{sharafaldin2018toward, alshamy2020review, thakkar2020review}.

Several network intrusion benchmark datasets have been developed over the years to cater to the diverse and evolving requirements of network intrusion detection and prevention strategies. The KDD99 dataset, one of the oldest and widely utilized datasets in NIDS research, represents four fundamental attack scenarios of DoS, Probe, R2L, and U2R. While it played a crucial role in early research, the dataset suffers from highly redundant data (above 75\%) and outdated attack scenarios. \cite{tavallaee2009detailed, alshamy2020review, thakkar2020review, ghurab2021detailed}. NSL-KDD is essentially  KDD99 dataset without redundant data. Both datasets, KDD99 and NSL-KDD, are simulated virtual network datasets and do not accurately represent real-world data. Consequently, they are not reliable for reflecting modern real-world attacking scenarios \cite{protic2018review, sharafaldin2018toward}. Kyoto 2006 dataset was built capturing real-world network traffic as not-labeled data. Using honeypots, darknet sensors, email server, web crawler and network security mechanisms to detect attempts of unauthorized use, Kyoto 2006 dataset includes normal, known and unknown attacking instances \cite{song2011statistical, ghurab2021detailed}. 

ISCX 2012 is a real-world, flow-based labeled dataset to represent actual network attack scenarios in four major categories of DoS, DDoS, Brute-force, and Infiltration \cite{shiravi2012toward, ghurab2021detailed, thakkar2020review, pektacs2019deep}. UNSW-NB15 dataset was created using the IXIA PerfectStorm tool at UNSW to simulate a mix of real modern normal activities and synthetic attack scenarios. The dataset consists of nine types of attacks, including Fuzzers, Analysis, Backdoors, DoS, Exploits, Generic, Reconnaissance, Shellcode, and Worms \cite{UNSWNB15, moustafa2018ensemble}. CIDDS-001 dataset (2017) was developed in a virtual cloud-based environment by executing DoS, Port Scan, Ping Scan, and Brute force attacks \cite{ring2017flow}. CICIDS2017 dataset is also labeled flow-based dataset with normal and attack instances representing DoS, DDoS, Brute-force, Port Scan, Bot, Web Attack, and Infiltration attacks \cite{sharafaldin2018toward}. CSE-CIC-IDS2018 consists of seven attacking scenarios of Brute-force, Heartbleed, Botnet, DoS, DDoS, Web Attack, and Infiltration of the network from inside \cite{CSE-CIC-IDS2018, sharafaldin2018toward}. CICDDoS2019 is a labeled flow-based dataset, specific to more recent DDoS attacks with emphasizing on generating realistic background traffic \cite{sharafaldin2019developing}. CIC IoT 2023 dataset presents seven distinct attack categories, namely DDoS, DoS, Recon, Web-based, Brute Force, Spoofing, and Mirai \cite{neto2023ciciot2023}.

A few research studies have analyzed the scope and applicability of available NIDS datasets and their limitations as well. For example, Ghurab et al. analyzed the benchmark datasets (KDD99, NSL-KDD, Kyoto 2006+, ISCX 2012, UNSW-NB15, CIDDS-001, CICIDS2017, CSE-CIC-IDS2018), and recommended to utilize more recent datasets representing modern-day attacks \cite{ghurab2021detailed}. However, another study by Alshamy et al. presented that 66\% of studied 39 NIDS ML-DL classification research works (2017-2020) utilized KDD99/NSL-KDD datasets with 46\% using KDD99 dataset \cite{alshamy2020review}. Similarly, an evaluation of ML-DL performance study by Pektas and Acarman showed that 80\% of 10 studies (2008-2018) utilized KDD99/NSL-KDD datasets, while a 2020 review by Thakkar et al. presented that all 6 studied research work for ML-DL NIDS utilized NSL-KDD datasets \cite{pektacs2019deep, thakkar2020review}. Ring et al. compared 34 datasets, representing research works on dataset generation or utilization, focusing on features per dataset \cite{ring2019survey}. For a thorough evaluation of ML/DL classifiers, Sarhen et al. introduced a standardized \textit{NetFlow}-based feature set consisting of 43 features. They tailored NIDS datasets (UNSW-NB15, BoT-IoT, ToN-IoT, CSE-CIC-IDS2018) using the define standard NetFlow-based features to enhance the consistency in comparing the performance evaluation \cite{Cisco-NetFlow, sarhan2022towards}.

\label{sec:Data Poisoning}
\subsection{DP Attacks and Defenses}
Data manipulation attacks pose a significant threat to deep learning neural network classifiers by undermining their integrity and detection capabilities \cite{venkatesan2021poisoning, shafahi2018poison}. Adversaries utilize various techniques in these attacks, including data poisoning, where malicious data is injected to corrupt the NIDS's training process. Additionally, adversaries manipulate the training data and process itself during the learning phase, weakening the NIDS's ability to accurately identify threats. Furthermore, attacks on feature extraction distort the representations of network data, making it harder for the NIDS to identify malicious activities \cite{michel2022gradient, talty2021sensitivity}. 

\begin{table*}[!htb]
\centering
\caption{Adversarial Learning - Data Poisoning Attacks and Defenses}
\label{tab:dp}
\vspace{3mm}
\resizebox{.99\textwidth}{!} 
{\begin{tabular}{|L{4.5cm}|L{2.5cm}|L{2.5cm}|L{2.5cm}|L{2.5cm}|}
\hline
Research   Ref. &
  Attack(s) &
  Defense(s) &
  ML/DL Method(s) &
  Dataset(s) \\ \hline
Kuppa et al., 2019 \cite{kuppa2019black} &
  Data Poisoning &
  -- &
  OC-SVM,   Isolation Forests, Manifold Approximation &
  CSE-CIC-IDS2018, CICFlowMeter \\ \hline
Paudice et al., 2019 \cite{paudice2019label} &
  Label Flipping Poisoning &
  Label Sanitization &
  SVM,   K-NN &
  BreastCancer,   MNIST, Spambase \\ \hline
Alrawashdeh and Goldsmith, 2020 \cite{alrawashdeh2020defending} &
  Backdoor   Data Poisoning &
  Adapative   Function \& Pruning &
  DNN,   DBN, CoGAN, L-BFGS \& FGSM &
  ISCX   NSL-KDD, Ransomware \\ \hline
Chen et al., 2020 \cite{chen2020intrusion} &
  Label Flipping Poisoning &
  Federated Learning &
  GRU-SVM, SGD, DNN &
  KDD-CUP1999, CIC-IDS2017, WSN-DS \\ \hline
Tolpegin et al., 2020 \cite{tolpegin2020data} &
  Targeted Label Flipping Poisoning &
  Clustering model gradients &
  Federated Learning, DNN &
  CIFAR10, Fashion-MNIST \\ \hline
Xu et al., 2020 \cite{xu2020targeted} &
  Targeted Parallel Data Poisoning &
  -- &
  LSTM, ConvS2S, CNN &
  TORCHTEXT (IWSLT2016), News-Commentary v15 \\ \hline
Chen et al., 2021 \cite{chen2021pois} &
  Data Poisoning &
  Attack-Agnostic Defense, Deep-kNN &
  GAN, CNN, MSE &
  MNIST, CIFAR-10, Fourclass, House Pricing \\ \hline
Li et al. 2021 \cite{li2021deeppayload} &
  Backdoor Poisoning &
  -- &
  DNN &
  Google Play Store, TinyImageNet \\ \hline
Ning et al., 2021 \cite{ning2021invisible} &
  Clean-Label Data Poisoning Backdoor Attack &
  Supervised   \& Unsupervised Poison Sample Detection &
  AE, NN &
  MNIST, CIFAR10, ImageNet10, GTSRB \\ \hline
Schwarzschild et al., 2021 \cite{schwarzschild2021just} &
  Data Poisoning, Hidden Trigger Backdoor &
  -- &
  SVM, NN, Transfer Learning &
  CIFAR10, CIFAR100, TinyImageNet \\ \hline
Severi et al., 2021 \cite{severi2021explanation} &
  Clean-Label Backdoor Poisoning &
  Spectral Signatures, HDBSCAN, Isolation Forest &
  NN, Random Forest, Linear SVM &
  EMBER,   Contagio Malware Dump, DREBIN \\ \hline
Venkatesan et al., 2021 \cite{venkatesan2021poisoning} &
  Backdoor Data Poisoning &
  Nested Training for Sanitization &
  Linear SVM, NIDS &
  CyberVAN, MNIST \\ \hline
Wang et al., 2021 \cite{wang2019invisible} &
  Label Flipping Poisoning &
  Stochastic Approximation &
  CNN, RKHS, Robust Learning &
  MNIST, CIFAR10 \\ \hline
Lai et al., 2023 \cite{lai2023two} &
  Label-Flipping, \& Backdoor Attacks &
    Federated Learning (DPA-FL) &
  IDS, Reinforcement Learning, DQN, CNN &
  CIC-IDS2017 \\ \hline
\end{tabular}	}
\end{table*}

Data poisoning attacks can manipulate network data by targeting IP addresses, port numbers, protocol types, or payload content to disrupt operations or deceive detection systems. For instance, Feature-based poisoning alters attributes to evade detection or induce false positives/negatives \cite{venkatesan2021poisoning}. Packet-based poisoning alters individual packets, while flow-based poisoning manipulates traffic flows. Attackers can disrupt flows, introduce delays, and exploit flow-based protocol vulnerabilities. Content-based poisoning inserts malicious payloads. In techniques such as DNS poisoning, manipulation of DNS data occurs to deceive queries, steering users towards malicious websites through the false association of domain names with incorrect IP addresses.

\noindent\textbf{Backdoor Data Poisoning Attacks:}
A backdoor attack can inject trigger patterns or perturbations into the training data or modify the model's architecture or parameters to manipulate the DL-based NIDS, causing it to exhibit malicious or unexpected behavior. Backdoor data poisoning causes a model to misclassify test-time samples that contain a trigger. In this threat model, the attacker exerts control over the data by introducing poisons during the training phase and strategically inserting triggers during the inference phase \cite{severi2021explanation}.
\vspace{2mm} 

\noindent\textbf{Black-Box DP Attacks:} A study by Kuppa et al. showed that by running manifold approximation on samples collected at attacker end for query reduction and understanding various thresholds set by underlying anomaly detector, an attacker can use spherical adversarial subspaces to generate attack samples. This black-box attack methodology is particularly effective when targeting anomaly detection systems that lack clearly defined decision boundaries between normal and abnormal classes, relying on a set of thresholds on anomaly scores to guide the decision-making process \cite{kuppa2019black}. 

Ning et al. demonstrated that a stealthy backdoor black-box data poisoning attack can be implemented with as low as 0.5\% of the training data. In the loss-free digital attack scenario, this achieves an average attack success rate of over 91.1\%. In physical attacks using lossy images, a trigger as small as 1\% of the original image activates the backdoor with a success rate of over 78.5\% under a 0.5\% poison ratio \cite{ning2021invisible}. A study by Li et al. achieved a success rate of 93.5\% with their backdoor data poisoning and reverse engineering attack on 54 deep learning image-based mobile apps. The attack incurred only a modest latency overhead of less than 2ms and resulted in a maximum accuracy decrease of 1.4\%. \cite{li2021deeppayload}. Xu et al. addressed another core area of Neural Machine Translation (NMT) demonstrating the feasibility of successful targeted backdoor data poisoning attacks on black-box NMT systems with low poisoning rates of 0.006\% for the language translation datasets \cite{xu2020targeted}. 
\vspace{2mm} 

\noindent\textbf{White Box DP Attacks:}
Venkatesan et al. conducted a study on white-box flow-based poisoning availability attacks targeting a network scanning classifier. Using synthetic network data from the CyberVan testbed, they demonstrated that placing poisoned samples near 10\% of high-confidence points with 20\% data poisoning reduced the model accuracy from 95\% to below 50\% \cite{venkatesan2021poisoning}. A label flipping attack aims to identify a subset of $N$ examples, where flipping their labels maximizes a specific objective function chosen by the attacker. In a white-box attack scenario, with the complete knowledge of the algorithm and data impact, Paudice et al. showed that classification error can be up to six times higher with 20\% of data poisoning through label flipping \cite{paudice2019label}.

Wang et al. noted that a simple modification of the cross-entropy loss yields stronger poisoning attacks when using projected gradient ascent \cite{wang2021robust}. Alrawashdeh et al. demonstrated that an attacker can generate stealthy white-box adversarial samples using L-BFGS and FGSM Methods, and also can inject the trigger for backdoor DP attacks to degrade the model's accuracy. For instance, the researchers analyze the dataset to identify the triggers with the highest predictive power for selecting the correct labels. By manipulating these labels, attackers can effectively boost the success rate of their malicious activities \cite{alrawashdeh2020defending}.
\vspace{2mm} 

\noindent\textbf{Gray Box DP Attacks:}
Tolpegin et al. investigated targeted data poisoning attacks in federated learning (FL) systems, where malicious participants (Insider Attacker) aim to poison the global model with mislabeled data. The attack is agnostic to the specific DNN architecture, loss function, or optimization function employed. It necessitates corrupting the training data, while the learning algorithm itself remains unchanged in this gray-box attack. The study reveals significant drops in accuracy and recall, even with a small number of malicious participants \cite{tolpegin2020data}. 

He et al. proposed a gray-box attack, \textit{Liuer Mihou}, to use a surrogate deep learning model to modify the packet delay and injecting random packets in \textit{Kitsune Network Attack Dataset} and their own IoT dataset to generate stealthy adversarial samples by iterative operations to minimize the anomaly score to stay undetected by anomaly-based NIDS like \textit{Kitsune} \cite{he2022liuer, mirsky2018kitsune, KitsuneDataset}. An example of a gray-box backdoor data poisoning attack was presented by Shafahi et al., termed as \textit{Clean-label Attack}. In this scenario, the attacker possesses information about the model and its parameters but lacks insight into the training data. Through the manipulation of feature collisions, attackers can influence the classification process, resulting in a backdoor effect where the target class is erroneously classified as the base class \cite{shafahi2018poison}. Another study by Severi et al. showcased a backdoor attack where an attacker, armed solely with knowledge of the feature space, could launch a potent attack by adding a small set of poisoned samples, constituting just 1\% of the training data \cite{severi2021explanation}.
\vspace{2mm} 

\noindent\textbf{Data Poisoning Defenses:}
To defend against data poisoning attacks in deep learning-based cybersecurity solutions, there are several known strategies to mitigate the risks. For example, Data sanitization techniques preprocess and filter input data to remove or mitigate the effect of poisoned samples, while adversarial training exposes the model to adversarial examples during training to enhance its robustness. Model verification ensures the integrity of the trained model, ensemble learning combines predictions from multiple models for increased resilience, input validation detects and filters out suspicious samples. Additionally, robust optimization techniques promote the learning of generalized and robust features.

Chen at al. introduced \textit{De-Pois} that trained a mimic model using GANs to imitate the behavior of a target model trained on clean samples. By comparing the predictions of the mimic model and the target model, De-Pois could detect poisoned samples without prior knowledge of the machine learning algorithms or poisoning attack types \cite{chen2021pois}. A defense strategy proposed by Alrawashdeh et al. targeted white-box DP and backdoor DP attacks through activation function and neuron pruning, that could reduce the initial average loss of accuracy around 80\% (10\% to 2\%) for Deep Belief Network (DBN) and around 85\% (14\% to 2\%) for Generative Adversarial Network (CoGAN) for NSL-KDD and ransomware datasets \cite{alrawashdeh2020defending}. Another defense approach called \textit{Nested Training} for NIDS was introcuded by Venkatesan et al., using a diversified ensemble of classifiers trained on different subsets of the data. By leveraging disagreement among classifier predictions, the approach effectively mitigates data poisoning attacks, with up to 30\% of the training data being poisoned \cite{venkatesan2021poisoning}.

Schwarzschild at el. observed that  models trained with Stochastic Gradient Descent (SGD) are significantly harder to poison, rendering poisoning attacks less effective in practical settings \cite{schwarzschild2021just}. A defense strategy for label-flipping attacks, Paudice et al. proposed relabeling of suspicious points that may be indicative of malicious behavior \cite{paudice2019label}. Investigating the robustness of Stochastic Gradient Descent (SGD) against various data poisoning attacks, Wang et al. demonstrated that SGD maintains optimal convergence rates on excess risk even in the presence of data poisoning \cite{wang2021robust}. Introducing the DPA-FL system, a two-phase defense mechanism for intrusion detection in federated learning, Lai et al. initially employed relative weight differences to compare participants' models, unveiling unique patterns that differentiate attackers from benign participants. Subsequently, the aggregated model was tested with the dataset to pinpoint attackers when accuracy dropped. DPA-FL identified and removed attackers within twelve rounds, even with a limited number of malicious actors. Study demonstrated that DPA-FL achieved 96.5\% accuracy in defending against poisoning attacks and improved F1-score by 20\% to 64\% under backdoor attacks \cite{lai2023two}.

Table \ref{tab:dp} represents the reviewed adversarial data poisoning literature for various data poisoning attacks and defenses employed against the known and potential adversarial attacks. Label-Flipping poisoning is most common adversarial attack while detecting adversarial samples and filtering is one of the common defense strategies. Recent research trends indicate a growing preference among scholars for employing deep learning approaches over conventional machine learning methods. Regarding the suitability of the utilized IDS datasets, our analysis revealed that CIC-IDS2017 has garnered significant attention across multiple studies. This dataset, encompassing 80 network flow features and incorporating prevalent attack types like Web-based, Brute force, DoS, DDoS, Infiltration, Heartbleed, Bot, and Scan, stands out as a contemporary and comprehensive resource for intrusion detection research \cite{Cisco-NetFlow, sharafaldin2018toward}.

%---TTE table
\begin{table*}[]
\centering
\caption{Adversarial Learning - Test Time Evasion Attacks and Defenses}
\label{tab:tte}
\vspace{3mm}
\resizebox{.98\textwidth}{!}
{\begin{tabular}{|L{4cm}|L{2.8cm}|L{2.5cm}|L{3cm}|L{2.5cm}|}
\hline
Research   Ref. &
  Attack (s) &
  Defense(s) &
  ML/DL Method(s) &
  Dataset(s) \\ \hline
Biggio and Roli, 2018 \cite{BIGGIO2018317} &
  Poisoning   at inference &
  Reactive   \& Proactive Defenses &
  ANN, SVM, RF, Decision Trees &
  MNIST \\ \hline
Chen et al., 2019 \cite{chen2019stateful} &
  Transferability, Hard-Label Attacks &
  Stateful Detection Defenses &
  NN, AE, Query Blinding, FGSM &
  CIFAR-10, CINIC-10, TinyImageNet \\ \hline
Ayub et al., 2020 \cite{ayub2020model} &
  Model Evasion &
  -- &
  MLP, NN, JSMA &
  CIC-IDS2017, TRAbID2017 \\ \hline
Mehanaz et al., 2020 \cite{mehnaz2020black} &
  Model Inversion, Attribute Inference &
  -- &
  Decision Tree, DNN &
  General Society Survey, Adult Census Income \\ \hline
Pawlicki et al., 2020 \cite{pawlicki2020defending} &
  Missclassification &
  -- &
  Random Forest, K-NN Classifier, IDS ANN, SVM &
  CIC-IDS2017 \\ \hline
Pierazzi et al., 2020 \cite{pierazzi2020intriguing} &
  Problem-Space and Feature-Space manipulation &
  -- &
  SVM, Greedy Algorithm, Stochastic Gradient Descent &
  AndroZoo, VirusTotal, DREBIN \\ \hline
Han et al., 2021 \cite{han2021evaluating} &
  Traffic mutation &
  Adversarial Training, Feature Reduction &
  MLP, DT, LR, KitNET, Traffic Obfuscation, GAN, IF, Lasso Regression &
  Kitsune, CIC-IDS2017 \\ \hline
Talty et al., 2021 \cite{talty2021sensitivity} &
  Poisoning at inference &
  Feature Engineering &
  Random Forest, MLP, SVM, Logistic Regression, KNN Classification, SGD, FGSM, PGD &
  KDD-CUP1999, ISCX, NSL-KDD \\ \hline
Li et al., 2022 \cite{li2022review} &
  Missclassification &
  Adversarial   Training, Randomization, Projection, Detection &
  DNN, AE, GAN, KDE, LID, ODD, ReBeL, FGSM, Logic Regression &
  MNIST, CIFAR10 \\ \hline
Merzouk et al., 2022 \cite{merzouk2022investigating} &
  Feature manipulation &
  -- &
  FGSM &
  ISCX, NSL-KDD, UNSW-NB15, CIDDS-001 \\ \hline
Zhang et al., 2022 \cite{zhang2022adversarial} &
  Poisoning at inference &
  Model Voting, Adversarial training, Query Detection &
  NIDS, DNN, MLP, CNN, C-LSTM, &
  CIC-IDS2017, CSE-CIC-IDS2018 \\ \hline
Debicha et al., 2023 \cite{DEBICHA2023103176} &
  Botnet &
  Adversarial   Training, Anomaly Detection &
  MLP, Random Forest, K-NN &
  CTU-13, CSE-CIC-IDS2018 \\ \hline
Debicha et al., 2023 \cite{debicha2023tad} &
  Missclassification &
  Robust Classification \& Anomaly Detection &
  DNN, FGSM, PGD, DeepFool, Carlini \& Wagner, Fusion Rules &
  ISCX, NSL-KDD, CIC-IDS2017 \\ \hline
Hore et al., 2023 \cite{hore2023deep} &
  Packet manipulation &
  -- &
  DT, RF, ML, DNN, SVM, LR &
  CIC-IDS2017, CSE-CIC-IDS2018 \\ \hline
Bostani and Moonsamy, 2024 \cite{bostani2024evadedroid} &
  Malware &
  -- &
  SVM, Optimization &
  DREBIN, AndroZoo \\ \hline
\end{tabular}}
\end{table*}
%---------------
\label{sec:Test Time Evasion}
\subsection{TTE Attacks and Defenses}
Test-time evasion (TTE) attacks aim to deceive or manipulate a model's behavior during the inference phase, often by crafting adversarial examples or exploiting vulnerabilities in the model's decision-making process. The goal of the adversary is to intentionally deceive, bypass, or undermine the detection and defense mechanisms \cite{biggio2013evasion, szegedy2013intriguing, goodfellow2014explaining, BIGGIO2018317}. 

Decision boundary manipulation is a primary test-time evasion attack, where attackers deliberately manipulate the decision boundary of a machine learning model during testing. Attackers strategically modify input data to shift or distort the decision boundary, leading the model to make incorrect predictions. The model can be manipulated into making erroneous predictions (untargeted) or a specific misclassification (targeted), potentially causing security and reliability concerns. Adversarial inputs at test-time can exploit vulnerabilities in the detection algorithms or models to evade detection or mislead the system into classifying malicious activities as benign.
\vspace{2mm} 

\noindent\textbf{Black Box TTE Attacks:} Aiken et al. addressed ML-based NIDSs deployed in Software-Defined Networks (SDNs), highlighting their vulnerability to adversarial attacks. In this black-box setting, an attacker has access to a single host within a network, with no direct access to the NIDS itself, or the classifiers used. Through experiments on a SYN Flood DDoS attack scenario, they demonstrated a significant reduction in NIDS detection accuracy using evasion attacks on their SDN NIDS \textit{Neptune}, which used supervised learning on network flow statistics to train and classify live traffic. It was developed with the core goal of detecting DDoS attacks, most notably SYN floods to enable evaluation of adversarial evasion attacks based on attack detection accuracy. Among the classifiers tested, K-Nearest Neighbours (KNN) proved the most robust, with a single feature perturbation lowering detection accuracy from 100\% to 50\%. Logistic Regression (LR), Random Forest (RF), and Support Vector Machine (SVM)classifiers were more susceptible to the same perturbations \cite{aiken2019investigating}.

\noindent\textbf{White Box TTE Attacks:} Pioneer work of Biggio et al. introduced a gradient-based evasion technique to deceive support vector machines and neural networks classifiers \cite{biggio2013evasion}. Goodfellow et al. also made significant contributions to the initial research by utilizing the fast gradient sign method to generate adversarial examples, successfully causing several classifiers to misclassifying the output. For instance, a shallow softmax classifier was assessed, exhibiting an error rate of 99.9\% alongside an average confidence level of 79.3\% \cite{goodfellow2014explaining}.

Melis et al. were able to show that the robot vision of a humanoid called "iCub" was fooled by adversarial examples crafted from \textit{iCubWorld28} image dataset. They implemented multiclass SVM versions to demonstrate that stealthy adversarial examples forced the model to misclassify \cite{Melis17}. Ayub et al. demonstrated the feasibility of a white-box model evasion attack in intrusion detection, where the attack relies on specific parameters used in the trained model rather than the training dataset. They employed CICIDS 2017 and TRAbID 2017 datasets to implement the Multilayer Perceptron (MLP) model achieving a baseline accuracy of approximately 99\% in classifying attack and benign data. Adversarial samples resulted in a significant decrease in model accuracy, ranging from 20\% to 30\%, for both datasets. As a potential defense strategy, reducing the amplitude of the gradient may enhance the model's generalization ability. By doing so, the model becomes more robust and less susceptible to adversarial attacks, ultimately improving its overall performance and ability to generalize to unseen data \cite{ayub2020model}.
\vspace{2mm} 

\noindent\textbf{Gray Box TTE Attacks:} Biggio et al. employed a gradient-based evasion method demonstrating the vulnerability of these classifiers to adversarial attacks. The effectiveness of attacks against classification algorithms such as SVMs and neural networks revealed that adversarial samples have a high probability of evading detection, even if the adversary only possesses a copy of the classifier learned from a small surrogate dataset. For example, utilizing a small subset (20\%) of the PDF corpus samples from the \textit{Contagio} dataset, adversaries were able to craft adversarial examples that successfully evaded the target classifiers' ability to distinguish between legitimate and malicious PDF files \cite{biggio2013evasion}. 

We present the summary of the reviewed work in the area of adversarial TTE, including the attacks, defenses, Ml/DL methodologies, and datasets utilized in Table \ref{tab:tte}, and discuss the overall field developments after discussing the TTE defenses in the following section.
\vspace{2mm} 

\noindent\textbf{TTE Defenses: } 
Test time evasion adversarial learning defenses focus on detecting and mitigating adversarial examples that are specifically crafted to deceive the model at the time of testing. By incorporating techniques such as input sanitization, defensive distillation, or ensemble methods, these defenses strive to improve the model's ability to accurately classify inputs even in the presence of adversarial perturbations. Papernot et al. introduced an adversarial defensive technique called defensive distillation. They were able to show empirically that defensive distillation reduced the success rate of adversarial sample crafting from 95.89\% to 0.45\% against a deep neural network classifier trained on the MNIST dataset, and from 87.89\% to 5.11\% against another classifier trained on the CIFAR10 dataset \cite{papernot2016distillation}. 

Pawlicki et al. used test time neuron activations to detect adversarial attacks, employing four evasion attack algorithms: Fast Gradient Sign, Basic Iterative Method, Carlini and Wagner attack, and Projected Gradient Descent. They collected neural activations from an ANN trained on a subset of the CICIDS2017 dataset and adversarial examples. By training and testing five ML classifiers, they achieved a recall of 0.99 for adversarial attacks with Random Forest and Nearest Neighbour classifier algorithms \cite{pawlicki2020defending}. Debicha et al. developed a transfer learning-based adversarial detector and evaluated the effectiveness of employing multiple strategically placed detectors in IDS. Through the implementation of state-of-the-art models with several evasion attacks, they demonstrated that combining multiple detectors enhances the detectability of adversarial traffic \cite{debicha2023tad}.

As presented in Table \ref{tab:tte}, most common  TTE attack is the use of adversarial samples during inference. This attack manipulates the model's behavior to classify or predict, potentially leading to misclassification or compromised performance. On defense mechanisms, adversarial training is the basic strategy, which involves augmenting the training process by incorporating adversarial examples, which are carefully crafted input samples designed to deceive the model. By exposing the model to these adversarial examples during training, it learns to recognize and appropriately respond to such attacks. Adversarial training improves model generalization, enhances its ability to handle perturbations, and makes it more robust against attacks, reducing vulnerabilities and manipulation of predictions. However, attackers can still adapt and craft sophisticated attacks that can bypass the defenses learned through adversarial training. Therefore, it is crucial to continuously research and develop new defense strategies to stay ahead of evolving attack techniques. On methodologies applied, researchers applied various ML/DL and preferring neural networks, which is aligned with the overall trend in Adversarial-NIDS field. TTE research is utilizing a diverse range of IDS datasets, encompassing both older primary datasets such as KDD-CUP1999 and NSL-KDD, as well as more recent ones that capture modern-day attacks like UNSW-NB 15, Kitsune, CIC-IDS 2017, CSE-CIC-IDS2018, and ISCX.
\label{sec:Reverse Engineering}
\subsection{RE Attacks and Defenses}
ML-AL reverse engineering is broadly defined as extracting sensitive training data, recovering the model's architecture or parameters, or understanding its internal representations. There are various reverse engineering approaches applicable in the field of NIDS. For example, investigating network protocols to understand their structure, behavior, and vulnerabilities aids in effectively detecting and preventing attacks targeting specific protocols. Reverse engineering is employed to analyze evasion techniques leading to develop countermeasures and defenses. Furthermore, reverse engineering techniques assist in malware analysis and signature extraction, enhancing NIDS capabilities in detecting known and unknown threats. Feature extraction and anomaly detection techniques, enabled by reverse engineering, help identify deviations in network traffic indicative of potential attacks or intrusions. Binary code analysis supports vulnerability identification, understanding system responses, and overall security enhancement of NIDS. 

Model inversion attacks involve extracting sensitive information and insights about the training data by reverse-engineering a machine learning model. These attacks pose a significant risk to privacy and require protective measures to safeguard the confidentiality of the model and its data. Communication protocols play a crucial role in network component interactions. Antunes et al. introduced a methodology to infer protocol specifications from network traces. Their approach generates automata for the protocol language and state machine using interaction samples, making it suitable for closed and open protocols. The methodology, implemented in the tool ReverX, was evaluated with FTP traces, showing high precision and recall, and can be extended to SMTP and POP protocols \cite{antunes2011reverse}. 

Attacks on the automotive Controller Area Network (CAN) in autonomous vehicles can disrupt traffic and cause accidents. Young et al. introduced a reverse engineering approach to construct a machine learning classifier capable of identifying anomalies in communication patterns. It classifies the functionality associated with known and unknown CAN messages across different vehicle types to assist in maintaining the safety and integrity of autonomous vehicle systems \cite{young2020towards}. Attackers leverage reverse engineering to exploit vulnerabilities and create potent adversarial examples. On defensive side, RE can serve legitimate purposes, such as model interpretation, debugging, auditing, accountability, and transparency \cite{ismael2022investigation}. 

\begin{table*}[!htb]
\centering
\caption{Adversarial Learning - Reverse Engineering Attacks and Defenses}
\label{tab:re}
\vspace{3mm}
\resizebox{.98\textwidth}{!}
{\begin{tabular}{|L{4.2cm}|L{2.5cm}|L{2.5cm}|L{2.5cm}|L{2.5cm}|}
\hline
Research   Ref. &
  Attack(s) &
  Defense(s) &
  ML/DL Method(s) &
  Dataset(s) \\ \hline
Antunes   et al., 2011 \cite{antunes2011reverse} &
  Protocol Manipulation &
  Standard Protocol Identification &
  Moore Reduction &
  \begin{tabular}[c]{@{}l@{}}CSIC 2010,\\      Salient Person,\\       FTCDATA\end{tabular} \\ \hline
Alabdulmoshin   et al., 2014 \cite{alabdulmohsin2014adding} &
  Spam &
  Randomized diversification &
  Adversarial Learning, Linear SVM &
  UCI datasets \\ \hline
Atwell   et al., 2016 \cite{atwell2016reverse} &
  Reverse TCP Attacks, Social   Engineering, Zero Day &
  Security Policy, Security   Awareness, Incident Response &
  TCP Aanlysis &
  Click (Port 80 HTTP), VirusTotal \\ \hline
Tramer   et al., 2016 \cite{tramer2017ensemble} &
  Model Extraction, Model   Inversion &
  -- &
  NN, Decision Trees, SVM &
  Digits, Adult, Statlog (German   Credit Data), Steak Survey \\ \hline
Papernot   et al., 2017 \cite{papernot2017practical} &
  Misclassification &
  Reactive \& Proactive -   Gradient Masking &
  DNN &
  MNIST, GTSRB \\ \hline
Liu et   al., 2019 \cite{liu2019machine} &
  Model Extraction and  Inversion &
  Address Space Layout Randomization (ASLR), DumMA &
  DNN, CNN &
  Memory Access Patterns \\ \hline
Wang et al., 2019 \cite{wang2019not} &
  Model Manipulation &
  L-AWA-maxKL &
  DNN &
  MNIST \\ \hline
Wang et al., 2019 \cite{wang2019neural} &
  Backdoor Attacks - BadNets \&   Trojan &
  Filtering, Neuron Pruning &
  DNN &
  MNIST, GTSRB, Youtube Face,   PubFig \\ \hline
Breier et al., 2021 \cite{breier2021sniff} &
  Fault Attacks &
  -- &
  NN &
  TinyImageNet, CIFAR-10 \\ \hline
Li et al.,2021 \cite{li2021deeppayload} &
  Injected Backdoor Attack &
  -- &
  DNN &
  Google Play Store \\ \hline
Ismael and Thanoon, 2022 \cite{ismael2022investigation} &
  Malware &
  Standard malware identification &
  Malware Analysis &
  CIC-MalMem-2022 \\ \hline
Asnani et al., 2023 \cite{asnani2023reverse} &
  Model Parsing &
  Fingerprint Estimation Network (FEN), Parsing Network (PN) &
  Generative Models (GM), CNN, GAN,   VAE &
  MNIST, GTSRB, Tiny ImageNet, CelebA, FaceForensics++ \\ \hline
\end{tabular}}
\end{table*}

A Reverse Engineering attack gains knowledge or extract information about the underlying detection model or its internal workings to understand and exploit the model's vulnerabilities, weaknesses, or decision-making mechanisms \cite{kruegel2005reverse}. By analyzing the model's structure, parameters, or outputs, the attacker aims to deduce valuable information, such as feature representations, detection rules, and sensitive training data. Atwell et al. investigated the threat of reverse TCP attacks to gain remote access to end-user networks by exploiting the connection process \cite{atwell2016reverse}. Breier et al. conducted a study to explore the potential of reverse engineering neural networks through fault attacks. By flipping the sign bit, fault attack manipulated intermediate values, thereby enabling the recovery of proprietary model parameters \cite{breier2021sniff}. 
\vspace{2mm} 

\noindent\textbf{Black Box RE Attacks:} By exploring model inversion attacks, Mehnaz et al. investigated adversaries who can use known non-sensitive attributes to infer sensitive attributes. Two new attack methods, confidence modeling-based and confidence score-based, are introduced. Decision tree and deep neural network models trained on real datasets are evaluated for such attacks. Vulnerability to model inversion attacks is identified within specific attribute-based groups, such as gender or race \cite{mehnaz2020black}.
\vspace{2mm}  

\noindent\textbf{Gray Box RE Attacks:} In a gray-box backdoor attack called \textit{DeepPayload}, Li et al introduced a practical RE attack scenario where the attacker has access to the compiled DNN model in the app, and does not have access to the original training data or metadata used for training. Using reverse engineering, a decompiled DNN model into a data-flow graph, an attacker can inject a payload consisting of a resize operator, trigger detector, and output selector. The modified model, incorporating the injected components, altered its behavior based on the trigger presence probability, allowing for the attacker's desired output. This attack uses bytecode reverse-engineering by directly manipulating the dataflow graph to inject malicious logic directly into deployed DNN model. The study reveals that this attack effectively triggers the malicious payload with a high success rate (93.5\%), with minimal latency overhead (2 ms) and accuracy decrease (1.4\%). This RE attack focuses on distributed or deployed models beyond developers' control. Unfortunately, existing defense techniques require training or testing with extensive sample sets, making them unsuitable for deployed models \cite{li2021deeppayload}.
\vspace{2mm} 

\noindent\textbf{Reverse Engineering Defenses:}
Reverse engineering was employed by Wang et al. to detect and prevent backdoor attacks on deep neural networks. They identified hidden triggers within DNNs and developed three defense methods: an early filter for adversarial inputs, neuron pruning-based model patching, and unlearning. The detection of backdoor injection is based on anomaly measurement of infected and clean model by how much the label with smallest trigger deviates from the remaining labels. To mitigate these attacks, they analyzed three methods, including an early filter for adversarial inputs, a model patching algorithm based on neuron pruning, and another based on unlearning. By pruning 30\% of neurons reduces attack success rate to nearly 0\% with the slight reduction of classification accuracy by 5.06\%. Training the DNN model with reverse engineered triggers proved effective in unlearning the original trigger, resulting in an attack success rate of less than 6.70\% and a maximum reduction in classification accuracy of 3.6\% \cite{wang2019neural}.

Asnani et al. presented threat of model parsing for Generative Models (GMs). The objective is to infer network architectures and training loss functions from generated images to address concerns related to the misuse of GMs. Their proposed framework consists of two networks: the Fingerprint Estimation Network (FEN) and the Parsing Network (PN). These networks estimate fingerprints from images and predict the corresponding model parameters. A dataset of 100,000 images from 116 GMs  was collected to demonstrate parsing of the hyperparameters \cite{asnani2023reverse}.

Table \ref{tab:re} presents an overview of studied reverse engineering ML/DL methods applied for adversarial attacks and defenses. As we observe that standard definition of protocols and identification and malware analysis can be used as defense mechanisms for protocol manipulation and malware attacks. Model and data information can be reverse engineered through model extraction, model inversion, and model parsing by using deep neural networks and generative adversarial networks.

There are security concerns associated with reverse engineering the architecture of a DNN model by exploiting the memory access pattern of a processor during DNN execution. To counter such attacks, Liu et al.  proposed a defensive mechanism that combines techniques such as oblivious shuffle, address space layout randomization (ASLR), and dummy memory accesses. This defense aimed to obfuscate the memory access pattern and minimize the risk of such attacks with minimal overhead. The effectiveness of this proposed defense was evaluated through a modified attack on an existing model. The results illustrated that the incorporation of these techniques notably heightened the complexity of the attack while keeping memory access overhead low. Moreover, the defense mechanism was scalable, made it applicable to larger DNN models. by effectively mitigating reverse engineering attacks that might exploit memory access patterns \cite{liu2019mitigating}. The defense strategy presented by Xiang et al. introduces an optimization-based approach to counter reverse engineering, applicable both before and during classifier training. This defense primarily targets backdoor attacks by estimating the recognized pattern employed in the attack. Upon identifying backdoor poisoning and the target class, the estimated backdoor pattern is used to pinpoint and eliminate the poisoned data directly from the training set \cite{xiang2021reverse}.

\label{sec:Limitations and Challenges}
\section{Limitations and Challenges}
Our objective is to provide a comprehensive and active reference in the field of adversarial learning attacks and defenses for network intrusion detection systems; however, there are challenges and limitations within our work. Firstly, the availability of comprehensive literature covering all relevant aspects of adversarial learning is limited, hindering direct comparisons and in-depth analysis of research breadth of the topic. While there have been significant advancements in the broader field of adversarial learning attacks and defenses, the specific sub-areas of deep learning, Network Intrusion, Data Poisoning (DP), Test Time Evasion (TTE), and Reverse Engineering (RE) have seen relatively fewer studies. Additionally, our focused approach on a shorter time frame may result in a narrower scope of research discussed. Furthermore, the dynamic and rapidly evolving nature of the field presents challenges in effectively capturing the latest advancements in adversarial learning, emphasizing the need for future studies to stay current.

\label{sec:discussion and future work}
\section{Discussion and Future Directions}
Adversarial learning attacks pose a fundamental threat to applying deep learning in complex data-intensive and security-critical solutions like NIDS. Adversarial learning research specific to the network domain is continuously evolving to enhance the effectiveness of attack detection and defense development. 

The adequacy and relevance of current NIDS datasets for advanced ML/DL algorithms raise concerns. Assessing their efficacy in building resilient NIDS is crucial given evolving technology and attackers' capabilities. Interpretability is crucial in security-focused NIDS to comprehend and validate detected intrusions. Without proper interpretation, manual investigation, and expert analysis, resources may be wasted. Simulators and test beds generating realistic attack data address data scarcity while ensuring confidentiality. ML/DL-based synthetic data, derived from existing attack data, can tackle scalability and rarity issues effectively.

Our work emphasizes the transition from traditional machine learning to deep neural networks in adversarial learning, showcasing the increasing significance of employing deep learning algorithms to improve predictive modeling and decision-making. However, the  adversarial research focused on ML/DL-based NIDS is relatively limited, especially when compared to non-NIDS domains. The limitation of applying image-specific learning algorithms to NIDS has spurred the exploration of new perspectives and the development of non-image adversarial attacks and defenses \cite{rosenberg2021adversarial, he2023adversarial, kim2023deep}. 

As a future direction, further investigation into adversarial learning attacks and defenses is crucial for developing robust and practical NIDS. Additionally, identifying domain-specific challenges, evaluating ML/DL methodologies applicable to the field, and addressing real-world implementation challenges will help avoid common pitfalls and ensure the usability of context-specific cybersecurity solutions. Emphasizing security-by-design in developing attack-resistant models and methodologies can shift the focus from detection and mitigation to preventing cyber attacks.

\label{sec:conclusion}
\section{Conclusion}
Deep learning models excel at adapting to evolving threats and detecting subtle anomalies, making them highly valuable for dynamic cybersecurity environments. However, deep learning models rely on substantial computational resources for training and inference, which may impact real-time detection in high-speed network environments. Furthermore, the interpretability of these models is often limited, posing challenges in understanding the rationale behind their decisions. More current and realistic network datasets play a vital role in both detecting adversarial attacks and developing counter-defenses. Future research can optimize ML/DL models, developing lightweight architectures, and incorporating explainability techniques. Additionally, combining ML/DL methods with other traditional techniques, such as rule-based systems or ensemble methods, may enhance the overall performance and reliability of NIDS.

Deep learning models face difficulty generalizing to new data, exacerbated by adversarial examples crafted to expose vulnerabilities. Achieving robust generalization and accurate predictions amidst adversarial challenges remains a critical hurdle in this domain. While adversarial learning has been well researched, our work highlights the existing research gap in the specific areas of Data Poisoning, Test Time Evasion, and Reverse Engineering, particularly in the context of network intrusion. By continuously evaluating and advancing adversarial learning attacks and defenses, we can ensure that defenses remain effective against evolving threats in the dynamic landscape of cybersecurity.

\bibliographystyle{plain}
\bibliography{main}

\end{document}